\documentclass[prl,twocolumn,english,showpacs]{revtex4-1}
\usepackage{amssymb}
\usepackage{color}
\usepackage{graphicx}
\usepackage{bbold}
\usepackage{bm}
\usepackage{amsmath}
\usepackage{amsfonts}
\usepackage{amssymb}
\usepackage{braket}
\begin{document}

\title{Measuring entropy and mutual information in the two-dimensional Hubbard model}

\author{E. Cocchi$^{1,2}$,  L. A. Miller$^{1,2}$, J. H. Drewes$^{1}$, C. F. Chan$^1$,  D. Pertot$^{1}$, F. Brennecke$^{1}$, and M. K{\"o}hl$^{1}$}

\affiliation{$^1$Physikalisches Institut, University of Bonn, Wegelerstrasse 8, 53115 Bonn, Germany\\
$^2$Cavendish Laboratory, University of Cambridge, JJ Thomson Avenue, Cambridge CB3 0HE, United Kingdom}

\begin{abstract}
We measure pressure and  entropy of ultracold fermionic atoms in an optical lattice for a range  of interaction strengths, temperatures and fillings. Our measurements demonstrate that, for low enough temperatures,  entropy-rich regions form locally in the metallic phase which are in contact with a Mott-insulating phase featuring lower entropy. In addition, we also measure the reduced density matrix of a single lattice site, and from the comparison between the local and  thermodynamic entropies we determine the mutual information between a single lattice site and the rest of the system. For low lattice fillings, we find the mutual information to be independent of interaction strength, however, for half filling we find that strong interactions suppress the correlations between a single site and the rest of the system.
\end{abstract}
\maketitle

Quantum mechanical correlations between particles give rise to collective behaviour beyond  intuitive imagination. Numerous classes of many-body states whose properties occur as result of quantum correlations are known to exist, such as Bose-Einstein condensates, Mott insulators, quantum magnets and superconductors. A general feature of a correlated  many-body system on a lattice is a strong correlation between  a single lattice site and  its surrounding environment. These correlations induce the sensitivity and vulnerability of a many-body state to external perturbations since even a very localized perturbation can destroy the nonlocal correlations.  Although the correlations are of microscopic origin, they are macroscopically manifest in the thermodynamic observables of the system. Reversing this argument, the correlations can be determined from precise thermodynamical measurements.

We explore the two-dimensional Hubbard model of spin-1/2 fermionic atoms in an optical lattice. The Hubbard model  considers the two elementary processes of tunneling between neighboring lattice sites with amplitude $t$ and on-site interaction between two fermions of opposite spin  with strength $U$. In a single-band approximation the Hubbard Hamiltonian reads
\begin{eqnarray}
\hat{H}=-t\sum_{\braket{i,j},\sigma}\left(\hat{c}_{i\sigma}^\dag \hat{c}_{j\sigma} +h.c.\right) +U\sum_{i}\hat{n}_{i\downarrow}\hat{n}_{i\uparrow}
\end{eqnarray}
Here $\hat{c}_{i\sigma}$ ($\hat{c}_{i\sigma}^\dag$) denotes the annihilation (creation) operator of a fermion on lattice site $i$ in spin state $\sigma=\{\uparrow, \downarrow\}$, the bracket $\braket{,}$ denotes the restricted sum over nearest neighbours, and $\hat{n}_{i\sigma}=\hat{c}^\dag_{i\sigma}\hat{c}_{i\sigma}$ is the number operator. A Mott insulator forms at half filling and strong repulsion, i.e. for $n= \braket{\hat{n}_{i\uparrow}}+\braket{\hat{n}_{i\downarrow}}=1$, and $U\gg t, k_BT$. It is characterized by an occupation of one particle per lattice site and an  energy gap for the creation of particle-hole excitations    of order $U$ \cite{Imada1998}. In contrast, for weak interactions and/or low lattice fillings, the fermions delocalize into Bloch waves and constitute a metallic state with finite charge compressibility.

To illustrate the thermodynamic implications of the microscopic physics we start from the atomic limit (single lattice site) with an average occupation of one particle, see Figure 1. For a high-temperature gas with $k_BT \gg U$, there are four equally probable microstates with zero or one fermion with spin-up or spin-down: $\ket{0_\uparrow,0_\downarrow}$, $\ket{0_\uparrow,1_\downarrow}$, $\ket{1_\uparrow,0_\downarrow}$ and $\ket{1_\uparrow,1_\downarrow}$. Hence, the entropy per site is $s=k_B \times 2\log(2)$. In contrast, at low-temperature, $k_BT \ll U$, particle number fluctuations such as double occupancies or empty lattice sites are suppressed by the charge excitation gap. The occupation per site is thus limited to the two cases $\ket{0_\uparrow,1_\downarrow}$ and $\ket{1_\uparrow,0_\downarrow}$ and  we expect $s=k_B\times \log(2)$. Furthermore, at high temperature, the entropy peaks at half filling whereas at low temperature exhibits a local maximum away from half filling, where it can increase above $s=k_B\times \log(2)$ due to the particle or hole doping. Hence, even in a very simple system, strong interactions can influence the thermodynamic observables.  In order to quantitatively understand the thermodynamic behaviour, we have to go  beyond the atomic limit and include tunnelling between nearest-neighbour lattice sites, thereby permitting the buildup of nonlocal density and spin correlations \cite{Greif2013,Hart2015,Parsons2016,Boll2016,Cheuk2016b,Drewes2016,Drewes2016b}.

In this work, we measure  pressure and entropy in the two-dimensional Hubbard model. From a comparison between the thermodynamic and the local (on-site) entropy, we determine the mutual information and thereby the amount of correlations between a single lattice site and its environment.  Previous measurements of the pressure and/or entropy in atomic Fermi gases have focussed on continuous (i.e. non-lattice) systems for non-interacting  \cite{Truscott2001} and  strongly-interacting  \cite{Nascimbene2010, Ku2012, Makhalov2014,Fenech2016,Martiyanov2016,Luo2007} Fermi gases. In a spin-polarized gas in an optical lattice, the entropy has been measured site-resolved in the atomic limit, i.e. disregarding fluctuations from tunneling \cite{Omran2015}. Our measurements extend beyond this by providing a spatially- (and thus filling-) resolved detection of the entropy without the zero-tunneling approximation. Finally, the mutual information between different subsystems of an optical lattice has been measured with bosonic atoms  \cite{Islam2015}.

\begin{figure}
 \includegraphics[width=.9\columnwidth,clip=true]{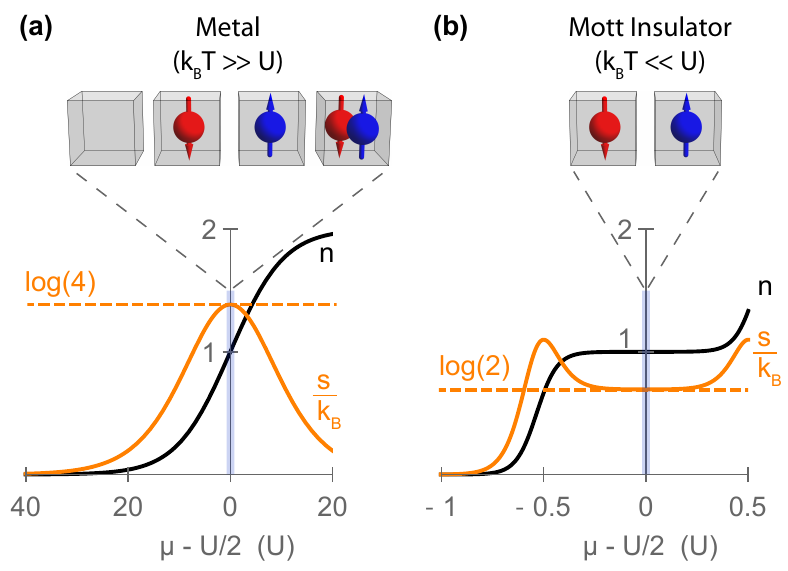}
 \caption{(Color online) Variation of lattice site occupation (black line) and entropy (orange line) with chemical potential in the atomic limit. (a) High-temperature gas with $k_BT \gg U$. For half filling, there are four equally probable microstates (see boxes) and hence the entropy peaks at $s=k_B\times \log(4)$. (b) In the low-temperature limit, $k_BT \ll U$,  doubles and holes are suppressed and hence the entropy is $s=k_B\times \log(2)$, however, it can raise above this value upon doping the system.}
\end{figure}

In our experiment, we prepare a spin-balanced  quantum degenerate mixture of the two lowest hyperfine states $\ket{F=9/2,m_F=-9/2}$ and $\ket{F=9/2,m_F=-7/2}$ of fermionic  $^{40}$K atoms \cite{Frohlich2011,Cocchi2016}. We load the quantum gas into an anisotropic, three-dimensional optical lattice in which  tunneling is suppressed along the vertical direction by means of a  high lattice depth. Hence, the dynamics is restricted to two-dimensional planes within which we choose a lattice depth of $5.2(1)\,E_\text{rec}\leq V_{xy}\leq 6.6(1)\,E_{\text{rec}}$, where $E_{\text{rec}}=\hbar^2 \pi^2/(2 m a^2)$ denotes the recoil energy, $a=532\,$nm is the lattice period, and $m$ is the atomic mass. The Hubbard interaction parameter $U$ is controlled by utilizing a Feshbach resonance near 202\,G which provides us with access to the parameter range from weak to strong interactions $0\lesssim U/t\lesssim 20$. Additionally, the temperature of the gas is adjusted by heating  due to a hold time in the optical lattice potential or periodic modulation of the trapping potential followed by a thermalization time.  We prepare equilibrium systems with well-defined parameters $t$, $U$, and $k_BT$ and detect the density distribution in a single two-dimensional layer of the optical lattice \cite{Cocchi2016}. By combining radio-frequency spectroscopy and absorption imaging we simultaneously detect the in-situ density distributions of singly-occupied lattice sites (``singles''),  $n_S=\braket{\hat{n}_{i\uparrow}-\hat{n}_{i\uparrow} \hat{n}_{i\downarrow}}$, and  doubly-occupied lattice sites (``doubles''), $n_D=\braket{\hat{n}_{i\uparrow} \hat{n}_{i\downarrow}}$. Our technique gives direct access to the equation of state  $n(\mu)$, where $\mu$ denotes the chemical potential. We perform thermometry by fitting the measured data with  numerical linked cluster expansion (NLCE) calculations of the two-dimensional Hubbard model \cite{Khatami2011} and the ideal Fermi gas on a square lattice.

\begin{figure}
 \includegraphics[width=\columnwidth,clip=true]{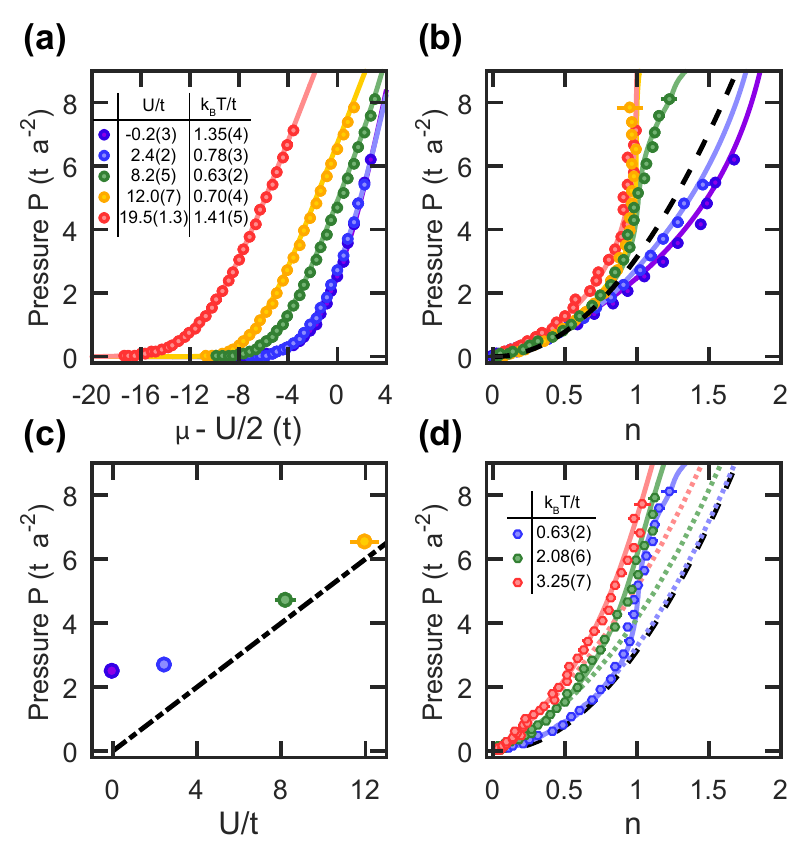}
 \caption{(Color online) Pressure as a function of interaction strength and temperature. (a) Pressure vs. chemical potential for different interaction strengths and temperatures.  (b) Pressure vs. filling for the same interactions and the same temperatures as in (a). (c) Pressure at half filling vs. interaction strength. The dash-dotted line is the infinite$-U$ and zero-temperature prediction $P=U/(2a^2)$. (d) Temperature dependence of pressure vs. filling for $U/t=8.2(5)$. The solid lines in (a,b,d) are the predictions from NLCE \cite{Khatami2011} with the exception of the purple solid line which represents the ideal Fermi gas on a lattice; the black, dashed line in (b,d) is the $T=0$ prediction of the ideal Fermi gas; the dotted lines in (d) are the predictions of the ideal Fermi gas at finite temperature.}
\label{fig2}
\end{figure}

We begin by determining the pressure from the measured equation of state \cite{Ho2010}. To this end, we start from the Gibbs-Duhem relation 
$S\,dT-A\,dP+N\,d\mu=0$, where $S$ denotes the entropy, $A$ the area, $P$ the pressure, and $N$ the total particle number. Expressing all extensive quantities in units per lattice site, pressure and density are related  to each other in thermal equilibrium and at constant temperature by 
\begin{equation}
P(\mu,T)=\frac{1}{a^2}\int_{-\infty}^\mu n(\mu^\prime,T) \,d\mu^\prime.
\end{equation}
In order to limit the noise in the numerical integration of the experimental data, we choose a lower bound of the integration region $\mu_\text{min}$ corresponding to a lattice site occupation of $n(\mu_\text{min})=0.01$. The resulting systematic uncertainty of the pressure is comparable to or  below the statistical uncertainty of our data. In Figure 2a, we show the measured pressure  as a function of the chemical potential at low temperature and different interaction strengths. The data are in excellent agreement with the predictions of NLCE theory, which we derive by numerical integration of the density \cite{Khatami2011}. In Figure 2b we show the same pressure data as a function of  $n$.  We find that, for low filling, experimentally determined pressures are nearly independent of interaction strength and agree well with the theoretical prediction of the ideal Fermi gas given by 
\begin{equation}
P_\text{2D}=-\frac{2k_BT}{\lambda_\text{dB}^2} \text{Li}_2(-\exp[- \mu/(k_B T)]).
\end{equation}
Here, $\lambda_\text{dB}$ is the thermal deBroglie wavelength computed using the effective mass at the bottom of the lattice band structure and $\text{Li}_2(x)$ is the polylogarithmic function of order 2. We attribute this behaviour to the suppression of interaction effects at low filling and the nearly-harmonic dispersion relation at the bottom of the band.  However, for $n \gtrsim 0.5$, we observe deviations from the ideal Fermi gas behaviour. For weak interactions, $U/t\lesssim 3$, the pressure is smaller than that of the ideal Fermi gas since for $n \gtrsim 0.5$ the particles experience the non-harmonic dispersion, which affects the pressure vs. density relation.  For strong interactions,  $U/t \gtrsim 8$, the pressure increases over that of the ideal Fermi gas and, in particular, develops a near-vertical slope at half filling when the lattice gas enters into a Mott insulator. This behaviour is associated with the opening of the charge gap of the Mott insulator, and one can understand the pressure at half filling in the limit of zero temperature and infinite interactions  by considering the internal energy $E=\braket{\hat{H}}=0$, which leads to $P=U/(2a^2)$. We plot this relation in Figure 2c and it asymptotically describes our data. In Figure 2d, we show the variation of pressure with increasing temperature for $U/t=8.2(5)$, compared to both the NLCE and ideal Fermi gas predictions. For low filling, the agreement with the prediction of the ideal gas is excellent and, in particular, one observes the change from the characteristic quadratic dependence $P \propto n^2$  of the low-temperature quantum gas to an approximately linear dependence $P \simeq n\cdot k_BT/a^2$ of a classical gas in the high-temperature limit. Near half filling, the vertical slope is washed out by  thermal excitations when $k_BT$ becomes of order $U$.

\begin{figure*}
 \includegraphics[width=.7\textwidth,clip=true]{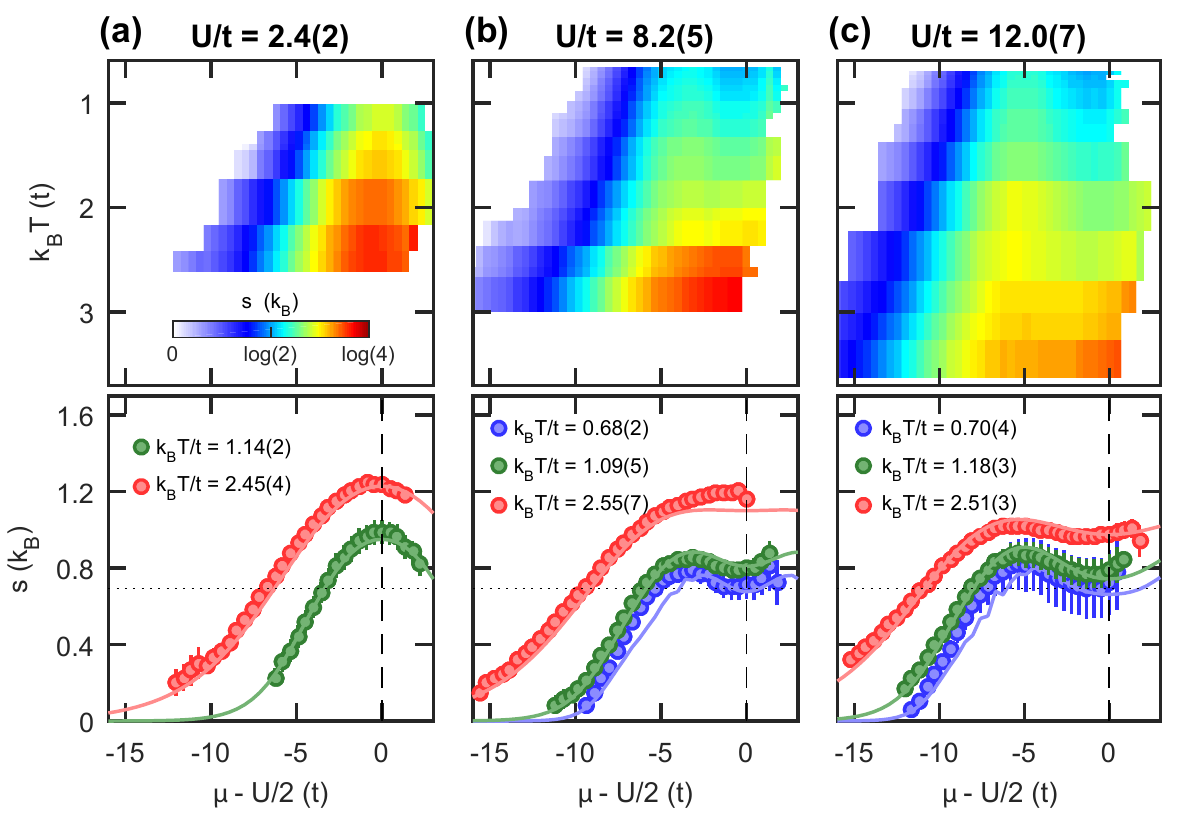}
 \caption{(Color online) Entropy vs. chemical potential and temperature.  The top row shows in color code the complete entropy data set and the bottom row shows entropy data at three selected temperatures together with the corresponding NLCE theory (solid lines).  For weak interactions  no Mott insulator forms and the entropy per site peaks at half filling at all temperatures explored. For intermediate and strong interactions at low temperature one observes a local minimum of the entropy per site owing to the charge excitation gap of the Mott insulator. }
\label{fig3}
\end{figure*}

We next determine the thermodynamic entropy from the measured pressure at constant chemical potential
\begin{equation}
s=a^2\frac{dP}{dT}\Big|_{\mu=\text{const.}}.
\end{equation}
In order to evaluate the entropy reliably, we have taken data sets very finely spaced in temperature increments of $k_B \Delta T \sim 0.2 t$. For each data set we determine the equation of state, compute the pressure, and then perform a numerical derivative with respect to temperature at a fixed chemical potential. In order to compute the numerical derivative, we interpolate the data in the temperature interval $[k_BT-t,k_BT+t]$ by a second-order polynomial and calculate the slope  at temperature $T$. A second-order polynomial was chosen as a minimal model to account for a non-constant heat capacity and at the same time minimise the number of fit parameters to yield a stable fit to the data. In Figure 3 we show the map of the measured entropy per site as a function of both temperature and chemical potential for three different interaction strengths $U/t=2.4, 8.2, 12$. For the weakest interaction, we do not observe a Mott insulator in the density profiles since the charge gap is washed out by the comparatively large kinetic energy. As a result, we observe the largest entropy per site at half filling for all temperatures (Fig. \ref{fig3}a). This is in agreement with the fact that for weak interactions the largest number of microstates is available at half filling. For strong interactions, $U/t \gtrsim 8$, and low temperatures, a Mott insulator forms at half filling, $\mu-U/2=0$,  surrounded by metallic phases for higher and lower chemical potential. We observe a non-monotonic variation of entropy vs. chemical potential with a local minimum at $\mu-U/2=0$ signalling that, at constant temperature, entropy is smaller in the gapped phase and higher in the thermally-connected gapless phase (Figs. \ref{fig3}b and c). By comparison of Figures \ref{fig3}b and \ref{fig3}c, we also show that for a larger gap this effect extends to higher temperatures, as expected. We attribute the deviations between experimental and NLCE data for the lowest temperatures near quarter filling to the 2nd-order polynomial fitting routine, which we have confirmed by analysing NLCE data with the same routine as the experimental data and comparing to the theoretically computed entropies.

\begin{figure}
 \includegraphics[width=\columnwidth,clip=true]{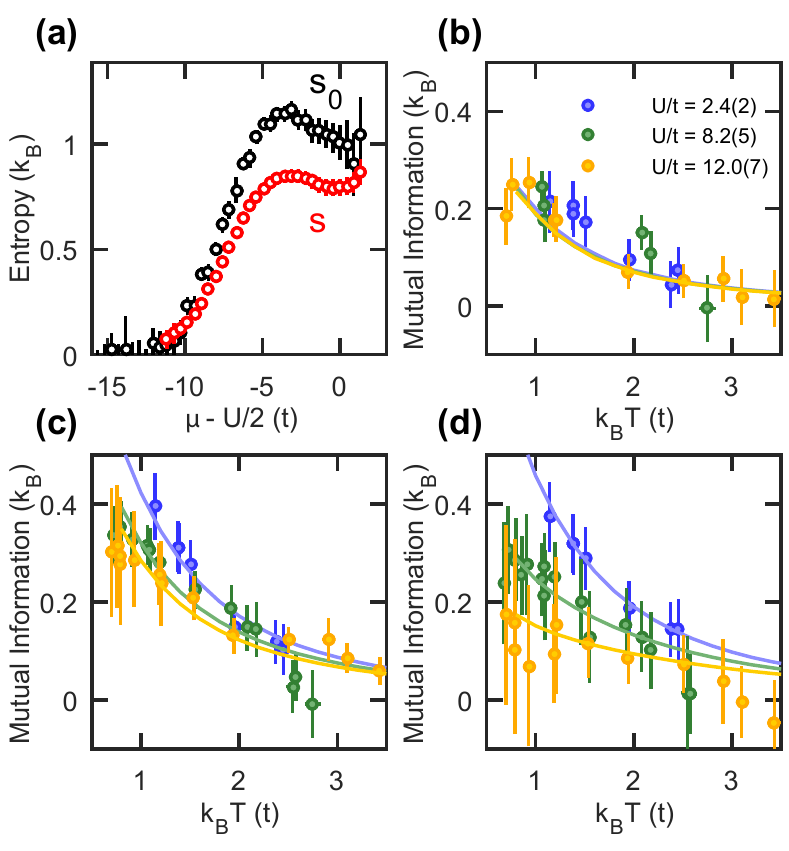}
 \caption{Mutual information between a single lattice site and its surrounding environment for different fillings, interactions and temperatures. (a) Thermodynamic entropy per site $s$ and on-site entropy $s_0$ for $U/t=8.2(5)$ and $k_BT/t=1.09(5)$. Mutual information for  $n=0.20(2)$ (b),  $n=0.67(2)$ (c), and $n=1.00(2)$ (d). The solid lines are theory predictions extracted from NLCE data. The color code is the same for all plots.}
\label{fig4}
\end{figure}

Finally, we turn our attention to the comparison between the thermodynamic and the local entropy which  quantifies the amount of correlations between a single lattice site and its environment. If one partitions a system into two subsystems $A$ and $B$ the amount of correlations between the two subsystems can be quantified by the mutual information $I=S_A+S_B-S_{AB}$ where $S_X = -k_B \text{Tr}  [\rho_X \log(\rho_X)]$ denotes the entropy of the reduced density matrix $\rho_X$ of subsystem $X=\{A, B\}$,  and $S_{AB}$ denotes the entropy of the full system.  In the following we consider the subsystem  $A$ to be a single lattice site, and subsystem $B$ to be the thermodynamic bulk excluding the single site $A$. The entropy $s_0$ of a single lattice site is directly determined from the single-site reduced density matrix by $s_0=-k_B\sum_i p_i \log(p_i)$. Here, $i=\{\uparrow\downarrow,\uparrow,\downarrow,0\}$ labels the probabilities $p_i$ for a site to be occupied with two particles, a spin-up particle, a spin-down particle or no particles, respectively. These are directly determined from the measured singles and doubles density distributions as $p_{\uparrow\downarrow}=n_D$, $p_\uparrow=p_\downarrow=n_S$ and $p_0=1-2n_S-n_D$ \cite{Zanardi2002}.  The entropy of the entire system with $L \gg 1$ sites is $S_{AB}=L\cdot s$, where $s$ is the measured thermodynamic entropy per site and, likewise, $S_B=(L-1)\cdot s$. Hence we obtain the mutual information as $I=s_0-s$, which we plot in Figure \ref{fig4} for various fillings, temperatures and interaction strengths. For low temperatures, we find a mutual information greater than zero, which indicates correlations between the site and its environment. We observe that for low filling, for which the effects of interactions are generally weak, the mutual information is independent of the interaction strength. In contrast, at half filling we observe larger correlations between the site and its environment for weak interactions than for strong interactions. This behaviour stems from the localization of atoms in the Mott insulator for strong interactions, whereas atoms are delocalized across the lattice for weak interactions. For high temperatures, the mutual information approaches zero indicating the absence of any correlations.

In conclusion, we have measured pressure and entropy distributions in the two-dimensional Hubbard model and have shown that at low temperatures a single lattice site develops correlations with the surrounding environment. The technique presented here determines the full thermodynamic entropy, including the entropy in the spin sector without the need for spin-resolved measurements. Hence it could find use in future attempts to cool strongly-correlated quantum gases by reshuffling the entropy \cite{Bernier2009,Ho2009}. 

We thank A. Daley and C. Kollath for discussions. The work has been supported by DFG (SFB/TR 185), the Alexander-von-Humboldt Stiftung, EPSRC and ERC (grant 616082).

\end{document}